\documentclass[twocolumn,pre,superscriptaddress]{revtex4}
\usepackage{graphics}
\usepackage{graphicx}
\usepackage{amsfonts}
\usepackage{textcomp}
\usepackage{amssymb}
\usepackage{mathrsfs}
\usepackage{amsmath}
\usepackage{color}
\usepackage{ulem}
\usepackage{float}

\begin{document}


\title{Disordered Hyperuniform Quasi-1D Materials}

\author{Duyu Chen\footnote{These authors contributed equally to this work.}}
\email[correspondence sent to: ]{duyu@alumni.princeton.edu}
\affiliation{Materials Research Laboratory, University of California, Santa Barbara, California 93106, United States}
\author{Yu Liu\footnotemark[1]}
\affiliation{HEDPS, Center for Applied Physics and Technology, School of Physics and College of Engineering, Peking University, Beijing 100871, People’s Republic of China} 
\author{Yu Zheng}
\affiliation{Department of Physics, Arizona State University, Tempe, AZ 85287}
\author{Houlong Zhuang}
\affiliation{Mechanical and Aerospace Engineering, Arizona State
University, Tempe, AZ 85287}
\author{Mohan Chen}
\affiliation{HEDPS, Center for Applied Physics and Technology, School of Physics and College of Engineering, Peking University, Beijing 100871, People’s Republic of China}
\author{Yang Jiao}
\affiliation{Materials Science and Engineering, Arizona State
University, Tempe, AZ 85287} \affiliation{Department of Physics,
Arizona State University, Tempe, AZ 85287}


\begin{abstract}
Carbon nanotubes are quasi-one-dimensional systems that possess superior transport, mechanical, optical, and chemical properties. In this work, we generalize the notion of disorder hyperuniformity, a recently discovered exotic state of matter with hidden long-range order, to quasi-one-dimensional materials. As a proof of concept, we then apply the generalized framework to quantify the density fluctuations in amorphous carbon nanotubes containing randomly distributed Stone-Wales defects. We demonstrate that all of these amorphous nanotubes are hyperuniform, i.e., the infinite-wavelength density fluctuations of these systems are completely suppressed, regardless of the diameter, rolling axis, number of rolling sheets, and defect fraction of the nanotubes. We find that these amorphous nanotubes are energetically more stable than nanotubes with periodically distributed Stone-Wales defects. Moreover, certain semiconducting defect-free carbon nanotubes become metallic as sufficiently large amounts of defects are randomly introduced. This structural study of amorphous nanotubes strengthens our fundamental understanding of these systems, and suggests possible exotic physical properties, as endowed by their disordered hyperuniformity. Our findings also shed light on the effect of dimensionality reduction on the hyperuniformity property of materials.




\end{abstract}
\maketitle

\section{Introduction}

Carbon nanotubes, a class of quasi-1D materials that can be conceptually constructed by rolling a graphene sheet, have been investigated extensively since their discovery \cite{91N-Iijima,93N-Iijima}. 
Benefiting from their superior physical, chemical, and mechanical properties, crystalline carbon nanotubes are popular candidates for a variety of applications, such as field-effect transistors \cite{98N-Tans}, rectifiers \cite{97S-Collins}, and sensors \cite{00S-Kong}.
However, the industrial production of carbon nanotubes (e.g., by catalytic chemical vapor deposition) is still not sufficiently well controlled and various defects can form during nanotube growth \cite{Zh03}. Thus, it is crucial to understand how the defects affect the physical properties and performance of carbon nanotubes.
For example, Robinson et al. demonstrated controlled introduction of oxidation
defects can enhance sensitivity of a SWNT network sensor to a variety of chemical vapors \cite{06NL-Robinson}.
Recently, Gifford et al. investigated the effect of the $sp^3$-hybridized defects on the energies of the optical emissive features and the influence of synthetic modifications on the resulting defect geometry toward attaining desired narrow photoluminescence capacity \cite{20ACR-Gifford}.



\begin{figure}[ht!]
\begin{center}
$\begin{array}{c}\\
\includegraphics[width=0.49\textwidth]{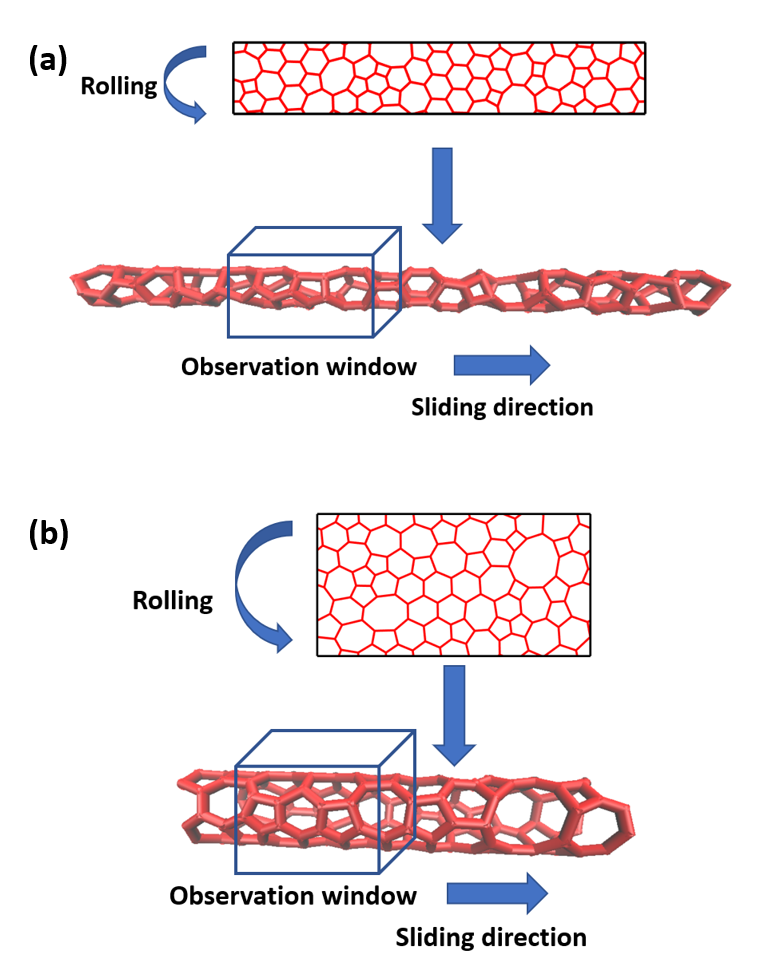} 
\end{array}$
\end{center}
\caption{(Color online) Illustration of a defected (3,0) zigzag (a) nanotube and a defected (3,3) armchair (b) nanotube formed by rolling graphene sheets with randomly distributed Stone-Wales defects along different rolling directions in two dimensions.} \label{fig_1}
\end{figure}

A variety of single-walled carbon nanotubes (SWNTs) can be formed by rolling up an infinitely long strip of a single graphene sheet along different directions. Conventionally, the type of SWNTs can be specified by a rolling vector $(n,m)$ (with $n>0$, $m\geq0$, and $n\geq m$) in the basis of two linearly independent vectors that connect a carbon atom in the graphene sheet to either two of its nearest atoms with the same bond directions \cite{Si01}. Two most common types of SWNTs are: (i) zigzag nanotubes with $n>0$ and $m=0$, and (ii) armchair nanotubes with $n=m$. Multi-walled nanotubes (MWNTs) consisting of purely zigzag nanotubes, or purely armchair nanotubes, or a mixture of both are also of great interest. 
A type of commonly seen defects in covalently-bonded network materials is the Stone-Wales (SW) topological defects \cite{St86}. A SW topological defect rotates a bond in the network by 90 degrees, transforming 4 adjacent hexagons into a pair of pentagons and a pair of heptagons in the absence of adjacent SW defects. Experimentally, the SW defects can be introduced into carbon nanotubes via high-energy radiations, during the synthesis of materials, or by applying strains \cite{Zh03}. In Fig. \ref{fig_1} we schematically shows the generation of a defected (3,0) zigzag nanotube and a defected (3,3) armchair nanotube with SW defects. Previous computational studies have also investigated the effect of the orientation and concentration of SW defects on the electronic properties of various single-walled carbon nanotubes at low defect concentrations \cite{Cr97, Az10, Pa13}. However, a comprehensive study of large-scale structures of carbon nanotubes with SW defects across a large range of defect concentrations is still lacking. 

Very recently, it is found that the SW defects in a variety of 2D material systems, including amorphous 2D silica \cite{Zh20}, amorphous graphene \cite{Ch21}, defected transition metal dichalcogenides \cite{PhysRevB.103.224102}, and defected pentagonal 2D materials \cite{Zh21}, can lead to special amorphous states \cite{Zh20, Ch21, Zh21, chen2021topological}, in which the density fluctuations in the materials are largely suppressed, a property termed ``hyperuniformity''. Interestingly, it was found that the disordered hyperuniform (DHU) states of these materials usually possess a significantly lower energy than other disorder models, suggesting the stability of DHU states for these systems over other disorder models. Moreover, DHU states can lead to unique electronic and thermal transport properties, resulted from mechanisms distinct from those that have been identified for their crystalline counterparts. A key discovery is that the SW defects in these systems play a significant role in establishing hyperuniformity of these 2D material systems \cite{Ch21}.

Disordered hyperuniformity (DHU) is a recently discovered exotic state of matter \cite{To03, To18a}, which lies between a perfect crystal and liquid. DHU systems are similar to liquids or glasses in that they are statistically isotropic and possess no Bragg peaks, and yet they completely suppress large-scale normalized density fluctuations like crystals. In this sense, DHU materials possess a hidden long-range order \cite{To03, Za09, To18a} similar to that in crystals. DHU is also manifested as the vanishing of static structure factor $S(k)$ in the infinite-wavelength (or zero-wavenumber) limit, i.e., 
\begin{equation}
\lim_{k\rightarrow 0}S(k) = 0, 
\label{eq_1}
\end{equation}
where $k$ is the wavenumber \cite{To03, To18a}. Here $S(k)$ is defined as  $S(k) \equiv 1 + \rho\Tilde{h}(k)$, where $\Tilde{h}(k)$ is the Fourier transform of the total correlation function $h(r) = g_2(r) - 1$, $g_2(r)$ is the pair correlation function, and $\rho$ is the number density of the system. Note that this definition implies that the forward scattering contribution to the diffraction pattern is omitted. For computational purposes, $S({k})$ is the angular-averaged version of $S({\bf k})$, which can be obtained directly from the particle positions ${\bf r}_j$, i.e.,
\begin{equation}
S({\bf k}) = \frac{1}{N} \left |{\sum_{j=1}^N \exp(i {\bf k} \cdot
{\bf r}_j)}\right |^2 \quad ({\bf k} \neq {\bf 0}),
\end{equation}
where $N$ is the total number of points in the system \cite{Za09}, and ${\bf k}$ is the wavevector. Equivalently, the local number variance $\sigma_N^2(R)\equiv \langle N^2(R)\rangle - \langle N(R) \rangle^2$ associated with a spherical observation window of radius $R$ grows more slowly than the window volume for DHU systems in the large-$R$ limit \cite{To03, To18a}, i.e., 
\begin{equation}
\lim_{R\rightarrow\infty} \frac{\sigma^2_N(R)}{R^d} = 0   
\label{eq_3}
\end{equation}
where $N(R)$ is the number of particles in a spherical window with radius $R$ randomly placed into the system, $R^d$ is proportional to the observation window volume, and $d$ the dimensionality of the system. The small-$k$ scaling behavior of $S(k) \sim k^\alpha$ dictates the large-$R$ asymptotic behavior of $\sigma_N^2(R)$, based on which all DHU systems can be categorized into three classes: $\sigma_N^2(R) \sim R^{d-1}$ for $\alpha>1$ (class I); $\sigma_N^2(R) \sim R^{d-1}\ln(R)$ for $\alpha=1$ (class II); and $\sigma_N^2(R) \sim R^{d-\alpha}$ for $0<\alpha<1$ (class III) \cite{To18a}.

Disordered hyperuniformity has been observed in a wide spectrum of physical and material systems \cite{To18a}, in both equilibrium and non-equilibrium, and both classical and quantum mechanical varieties. Examples include the density fluctuations in early
universe \cite{Ga02}, maximally random jammed packing of hard particles
\cite{Za11a, Ji14, Ch14}, exotic classical ground
states of many-body systems \cite{Za11b, To15}, jammed colloidal systems \cite{Ku11, Dr15}, driven non-equilibrium systems \cite{Ja15, We15}, certain quantum ground states \cite{To08, Fe56}, avian photoreceptor patterns \cite{Ji14}, organization of immune systems \cite{Ma15}, amorphous silicon
\cite{He13, Xi13}, a wide class of disordered cellular materials
\cite{Kl19}, dynamic random organizing systems
\cite{hexner2017noise, hexner2017enhanced, weijs2017mixing,
Le19a, Le19b}, electron density distributions \cite{Ge19, sakai2022quantum}, and vortex distribution in superconductors \cite{Ru19, Sa19}, to name but a few. 

Since nanotubes can be constructed by rolling the corresponding sheets of 2D materials, it is natural to ask the question: Are these quasi-1D materials containing Stone-Wales topological defects are also hyperuniform? In this paper, we address this question by generalizing the notion of hyperuniformity to quasi-1D materials, which can be considered as 1D projections of higher-dimensional structures in this context, and may involve the non-trivial situations where multiple points in the higher-dimensional structures are mapped to the same point in the projection. In particular, we systematically generate a diverse spectrum of amorphous carbon nanotubes by continuously introducing Stone-Wales (SW) topological defects into the crystalline systems, and quantify the generalized density fluctuations in these systems. We demonstrate that all amorphous nanotubes are hyperuniform, i.e., the infinite-wavelength density fluctuations of these systems are completely suppressed, and the systems possess hidden long-range order, regardless of the diameter, rolling axis, number of rolling sheets, and defect fraction of the nanotubes. This structural study of amorphous nanotubes strengthens our fundamental understanding of these systems, and suggests possible exotic physical properties, as endowed by their disordered hyperuniformity.







The rest of the paper is organized as follows: In Sec. II we discuss the the generalization of the hyperuniformity concept to quasi-1D materials. In Sec. III we discuss the procedures to generate various types of carbon nanotubes with randomly distributed SW defects. In Secs. IV and V, we respectively investigate the density fluctuations of single-walled and multi-walled nanotubes, which we use as examples to demonstrate the use of the theoretical framework for analyzing hyperuniformity properties in quasi-1D materials. In Sec. VI, we report the physical properties of the hyperuniform carbon nanotubes, including their stability and density of states. In Sec. VII, we provide concluding remarks and discuss the implications of our work in other quasi-1D material systems. 

\section{Hyperuniformity of quasi-1D materials}

In this section we introduce the generalization of the hyperuniformity concept to quasi-1D materials. These systems typically have small width in all the directions other than the axial/propagation direction. For example, carbon nanotubes have small finite width in the rolling direction. Since hyperuniformity is a large-scale structural feature, only the density fluctuations along the propagation direction should be relevant in the context of hyperuniformity, and effectively we are looking at 1D projections of the graphene sheets along the axial/propagation direction in the case of carbon nanotubes.

An important issue that we need to address when looking at density fluctuations of low-dimensional projections of higher-dimensional structures is that multiple points in the higher-dimensional structures can be mapped to the same point in the projections. Here we generalize the definition of the structure factor $S({\bf k})$ to be:
\begin{equation}
    S({\bf k}) = \frac{1}{G} \left |{\sum_{j=1}^M g_j\exp(i {\bf k} \cdot
{\bf x}_j)}\right |^2 \quad ({\bf k} \neq {\bf 0}),
\end{equation}
where $M$ is the number of distinguishable points in the projections, the {\it multiplicity} $g_j$ is defined as the number of points in the higher-dimensional structures that are mapped to the given point ${\bf x}_j$ in the projections, and $G=\sum_{j=1}^M g_j$. Accordingly, we generate the concept of $N(R)\equiv \langle N(R;{\bf x}_0)\rangle$ associated with $\sigma_N^2(R)$ to be:
\begin{equation}
    N(R;{\bf x}_0) = \sum_{j=1}^M g_j m({\bf x}_j-{\bf x}_0; R),
\end{equation}
where $\langle \cdots \rangle$ denotes ensemble average, $m({\bf x}-{\bf x}_0; R)$ is the indicator function of the observation window centered at ${\bf x}_0$ with radius $R$ and is defined as
\begin{equation}
m({\bf x; R}) = \left \{
\begin{array}{c@{\hspace{0.3cm}}c@{\hspace{0.3cm}}c}
1, & |{\bf x}| \leq R, &
\\ 0, & \textnormal{otherwise}. & \end{array} \right .
\end{equation}
We note that using 1D observation window to look at $\sigma_N^2(R)$ of the projection is equivalent to using observation window in the higher dimension that encompasses the quasi-1D materials in the radial directions, but slide in the axial/propagation direction, which is illustrated in Fig. \ref{fig_1}. With these generalizations of $S(k)$ and $\sigma_N^2(R)$, we can then use the definitions of hyperuniformity specfied in Eqs. \ref{eq_1} and \ref{eq_3} for a low-dimensional projection, and use the dimension of the projection for $d$ in Eq. \ref{eq_3}.

\section{Generation of amorphous carbon nanotubes containing Stone-Wales defects}

To generate a defected zigzag nanotube at a given defect concentration $p$, we first introduce prescribed number of randomly distributed SW defects $N_d = N_b p = \frac{3}{2}Np$ into a graphene sheet with finite width $L_y$ in the vertical direction in Fig. \ref{fig_1}(a), where $N_b = \frac{3}{2} N$ and $N$ are the number of bonds and atoms in the sheet, respectively, and $L_y$ is the length of the $(n,0)$ vector. We then relax the structure according to the procedure described in Ref. \cite{Ch21}, and subsequently roll up the sheet along the vertical direction shown in Fig. \ref{fig_1}(a). Specifically, for a given atom at location $(x,y)$ in a graphene sheet, its coordinate $(x',y',z')$ in the resulting zigzag nanotube is given by 
\begin{equation}
    \left \{
\begin{array}{l}
x'= \frac{L_y}{2\pi}\cos(\frac{2\pi y}{L_y}),  
\\  y'= \frac{L_y}{2\pi}\sin(\frac{2\pi y}{L_y}), 
\\ z'= x  \end{array} \right.
\end{equation}
where $x$ and $y$ axis are the horizontal and vertical directions in Fig. \ref{fig_1}(a). On the other hand, to generate a defected armchair nanotube at a given defect concentration $p$, we first follow similar procedure described in Ref. \cite{Ch21}, but use a graphene sheet with finite width $L_x$ in the horizontal direction in Fig. \ref{fig_1}(a), where $L_x$ is the length of the $(n,n)$ vector. We note that the directions in Fig. \ref{fig_1}(b) are rotated by 90 degrees from those in Fig. \ref{fig_1}(a) for clear visualization, so the horizontal direction in Fig. \ref{fig_1}(a) is the vertical direction in Fig. \ref{fig_1}(b). We then roll up the sheet along the horizontal direction shown in Fig. \ref{fig_1}(a), or the vertical direction in Fig. \ref{fig_1}(b). Specifically, for a given atom at location $(x,y)$ in a graphene sheet, its coordinate $(x',y',z')$ in the resulting armchair nanotube is given by 
\begin{equation}
    \left \{
\begin{array}{l}
x'= \frac{L_x}{2\pi}\cos(\frac{2\pi x}{L_x}),  
\\  y'= \frac{L_x}{2\pi}\sin(\frac{2\pi x}{L_x}), 
\\ z'= y  \end{array} \right.
\end{equation}
where $x$ and $y$ axis are the horizontal and vertical directions in Fig. \ref{fig_1}(a).

To obtain multi-walled carbon nanotubes with SW defects consisting of concentric single-walled nanotubes, we simply generate the constituting single-walled nanotubes separately according to the procedures above, and combine the atoms together. This is possible because the aforementioned procedure produces nanotubes that all wrap around the $z'$ axis in three dimensions.

\section{Density fluctuations of single-walled nanotubes}
To investigate density fluctuations of zigzag and armchair single-walled nanotubes, we employ the procedures described in Sec. III to generate 10 configurations of (3,0), (5,0), (3,3), and (5,5) nanotubes at each of the given defect concentrations $p=0.02$, $0.04$, $0.06$, $0.08$, $0.10$, $0.12$, and $0.14$. We then compute their local number variance $\sigma_{N}^2(R)$ and structure factor $S(k)$. This setting also allows us to study the effect of nanotube radius on the structural features of nanotubes.

As a reference, it can be easily seen that defect-free $(n,0)$-zigzag nanotubes are mapped into two-scale 1D point patterns with $\zeta = \frac{1}{3}$ \cite{To03} when projecting onto the cylinder axis of the nanotubes, with $n$ carbon atoms superimposed onto each other at each point in the 1D projected point pattern. The number variance of the $(n,0)$-zigzag nanotubes $\sigma_{N}^2(R)$ can then be determined analytically as 
\begin{equation}
\sigma_{N}^2(R) = n^2 \sigma_{N,D}^2(R),
\end{equation}
where $\sigma_{N,D}^2(R)$ is the number variance of the projected two-scale 1D point patterns with $\zeta = \frac{1}{3}$ \cite{To03}. As a result, $\sigma_{N}^2(R)$ of defect-free zigzag nanotubes fluctuate around certain constant, whose value depends on $n$, and these nanotubes are class-I hyperuniform.

\begin{figure*}[ht!]
\begin{center}
$\begin{array}{c}\\
\includegraphics[width=0.985\textwidth]{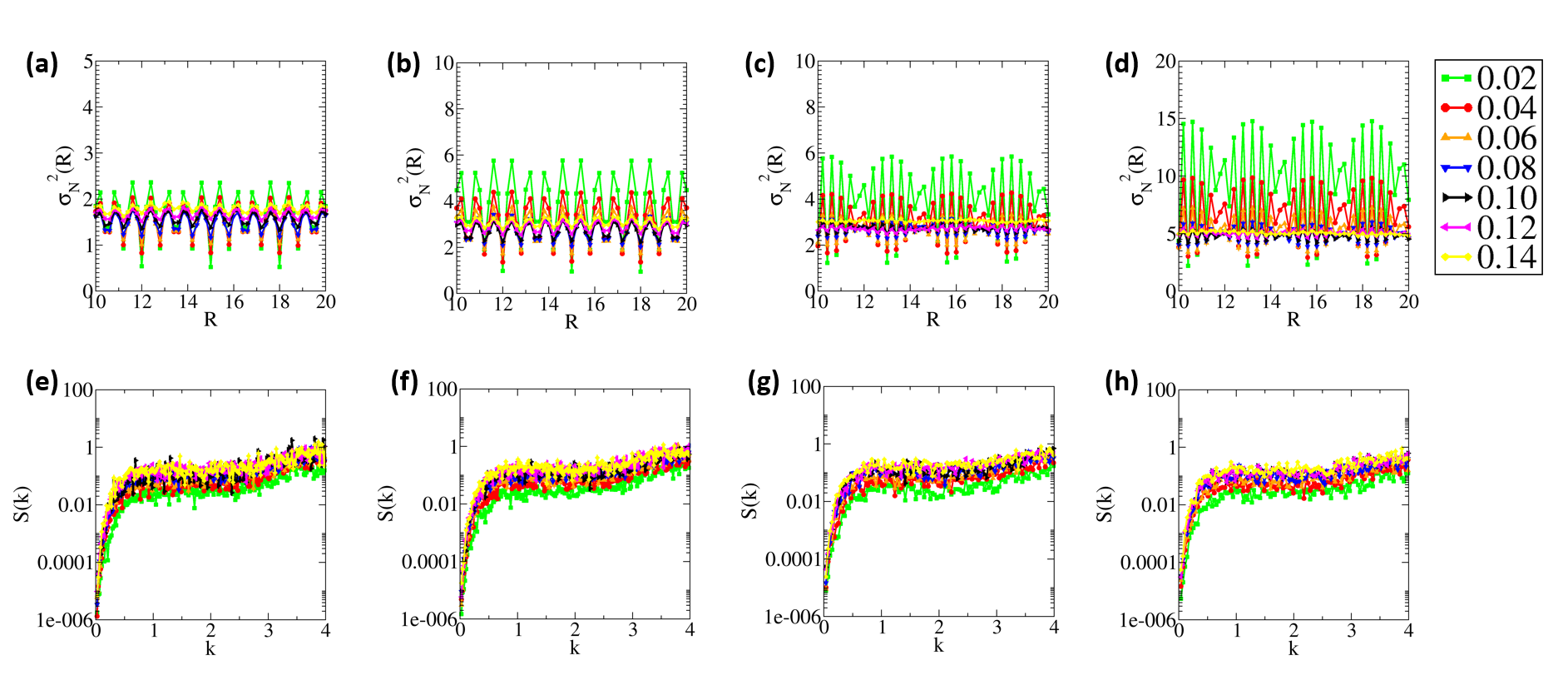} 
\end{array}$
\end{center}
\caption{(Color online) Local number variances $\sigma_N^2(R)$ (Top) and structure factor $S(k)$ (Bottom) of defected single-walled zigzag and armchair nanotubes at different defect concentrations $p$. (Column 1) (3,0) zigzag. (Column 2) (5,0) zigzag. (Column 3) (3,3) armchair. (d) (Column 4) armchair. The results are all averaged over 10 configurations.} \label{fig_2}
\end{figure*}

As SW defects are introduced into the nanotubes, the structures gradually transition into amorphous ones, which are reflected in their local number variance $\sigma_{N}^2(R)$. For example, we show the computed $\sigma_{N}^2(R)$ for $(3,0)$-zigzag nanotubes at different defect fractions $p$ in Fig. \ref{fig_2}(a). At low $p$, $\sigma_{N}^2(R)$ exhibits ``periodicity'' in the window radius $R$, suggesting the reminiscence of the crystalline order in the systems; at large $p$, $\sigma_{N}^2(R)$ varies smoothly with damped oscillations as $R$ increases, indicating the emergence of truly amorphous states. Also, by comparing the results of (3,0)-zigzag nanotubes in Fig. 2\ref{fig_2}(a) and the results of (5,0)-zigzag nanotubes in Fig. \ref{fig_2}(b), one can see that at given $p$, increasing the radius (or decreasing the curvature) of the armchair nanotube increases density fluctuations. Interestingly, the variance $\sigma_{N}^2(R)$ of these zigzag nanotubes fluctuate around certain constants as $R$ increases at all investigated $p$, indicating that these structures are class-I hyperuniform. This is related to the fact that SW transformation and subsequent structural relaxations only result in local rearrangements of atoms, and do not affect the scaling of large-scale density fluctuations in these nanotubes.

Similarly, defect-free $(n,n)$-armchair nanotubes are mapped into single-scale 1D point patterns \cite{To03} when projecting onto the cylinder axis of the nanotubes, with $2n$ carbon atoms superimposed onto each other at each point in the 1D point pattern. The number variance of the $(n,n)$-armchair nanotubes $\sigma_{N}^2(R)$ can then be determined analytically as 
\begin{equation}
\sigma_{N}^2(R) = 4n^2 \sigma_{N,S}^2(R), 
\end{equation}
where $\sigma_{N,S}^2(R)$ is the number variance of the projected single-scale 1D pattern. As a result, $\sigma_{N}^2(R)$ of defect-free armchair nanotubes fluctuate around certain constant, whose value depends on $n$, and these nanotubes are class-I hyperuniform.

As SW defects are introduced into the nanotubes, the behaviors of these armchair nanotubes are similar to those of the zigzag nanotubes. For example, we show the computed $\sigma_{N}^2(R)$ for $(3,3)$-armchair nanotubes at different defect fractions $p$ in Fig. \ref{fig_2}(c). Importantly, at all investigated $p$, the variance $\sigma_{N}^2(R)$ of these armchair nanotubes fluctuate around certain constants, indicating that these structures are hyperuniform. Also, by comparing the results of (5,0)-zigzag nanotubes in Fig. \ref{fig_2}(c) and the results of (5,5)-zigzag nanotubes in Fig. \ref{fig_2}(d), one can see that at given $p$, increasing the radius (or decreasing the curvature) of the armchair nanotube increases density fluctuations.

The results of ensemble-averaged $S(k)$ of the aforementioned zigzag and armchair nanotubes are shown in Fig. \ref{fig_2}(e)-(h), which all decreases to zero as $k$ goes to zero, regardless of defect concentration $p$. These results further confirm the hyperuniformity of these nanotubes, and are consistent with the results of $\sigma_{N}^2(R)$.

\section{Density fluctuations of multi-walled nanotubes}
It can be easily seen that MWNTs consisting of $K$ defect-free zigzag nanotubes with $(n_1,0)$, $(n_2,0)$, $\cdots$, $(n_K,0)$ rolling vectors, respectively, are mapped into two-scale 1D point patterns with $\zeta = \frac{1}{3}$ \cite{To03} when projecting onto the cylinder axis of the nanotubes, with $n_1+n_2+\cdots+n_K$ carbon atoms superimposed onto each other at each point in the 1D point pattern. The number variance of the MWNTs $\sigma_{N}^2(R)$ can then be determined analytically as
\begin{equation}
\sigma_{N}^2(R) = (n_1+n_2+\cdots+n_K)^2 \sigma_{N,D}^2(R), 
\end{equation}
where $\sigma_{N,D}^2(R)$ is the number variance of the projected two-scale 1D point patterns with $\zeta = \frac{1}{3}$ \cite{To03}. Therefore, $\sigma_{N}^2(R)$ of defect-free MWNTs consisting of multiple zigzag nanotubes fluctuate around certain constant, and these nanotubes are class-I hyperuniform.

\begin{figure*}[ht!]
\begin{center}
$\begin{array}{c}\\
\includegraphics[width=0.745\textwidth]{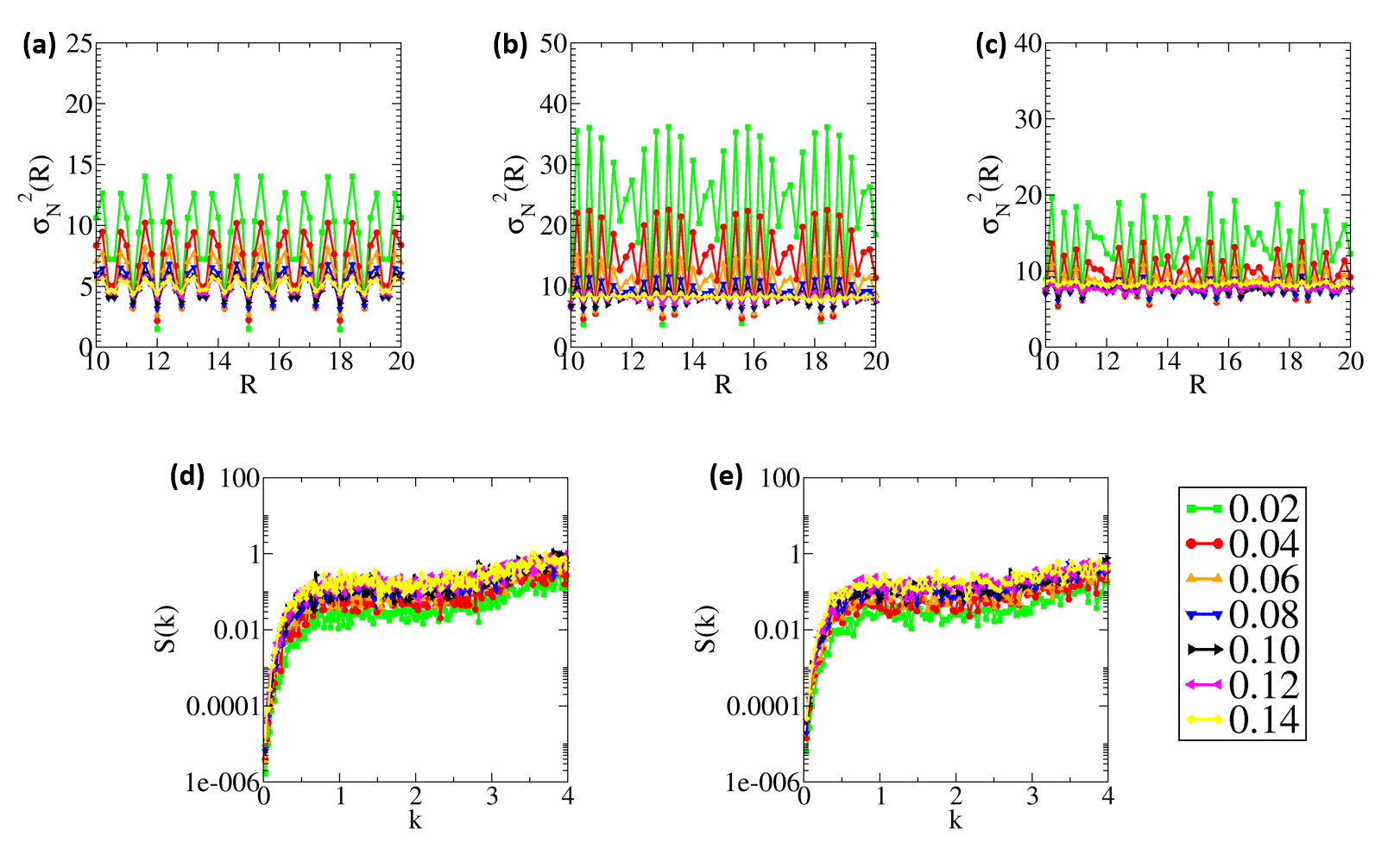} 
\end{array}$
\end{center}
\caption{(Color online) Local number variances $\sigma_N^2(R)$ and structure factor $S(k)$ of defected multi-walled nanotubes at different defect concentrations $p$. (a) $\sigma_N^2(R)$ of double-walled nanotubes consisting of a (3,0) zigzag nanotube and a (5,0) zigzag nanotube. (b) $\sigma_N^2(R)$ of double-walled nanotubes consisting of a (3,3) armchair nanotube and a (5,5) armchair nanotube. (c) $\sigma_N^2(R)$ of double-walled nanotubes consisting of a (5,0) zigzag nanotube and a (5,5) armchair nanotube. (d) $S(k)$ of double-walled nanotubes consisting of a (3,0) zigzag nanotube and a (5,0) zigzag nanotube. (e) $S(k)$ of double-walled nanotubes consisting of a (3,3) armchair nanotube and a (5,5) armchair nanotube. The results are all averaged over 10 configurations.} \label{fig_3}
\end{figure*}

On the other hand, as SW defects are introduced into the nanotubes, the structures gradually transition into amorphous ones, which are reflected in their local number variance $\sigma_{N}^2(R)$. For example, we show the ensemble-averaged $\sigma_{N}^2(R)$ for MWNTs consisting of a $(3,0)$-zigzag nanotube and a $(5,0)$-zigzag nanotube at different defect fractions $p$ in Fig. \ref{fig_3}(a). At low $p$, $\sigma_{N}^2(R)$ exhibits ``periodicity'' in the window radius $R$, indicating that the crystalline order is reminiscent in the systems; at large $p$, the oscillations of $\sigma_{N}^2(R)$ become much more damped as $R$ increases, suggesting the emergence of truly amorphous states. Interestingly, the variance $\sigma_{N}^2(R)$ of these MWNTs fluctuate around certain constants as $R$ increases at all investigated $p$, indicating that these structures are class-I hyperuniform. These behaviors are similar to those in the cases of the SWNTs.

Similarly, MWNTs consisting of $K$ defect-free armchair nanotubes with $(n_1,n_1)$, $(n_2,n_2)$, $\cdots$, $(n_K,n_K)$ rolling vectors, respectively, are mapped into are mapped into single-scale 1D point patterns \cite{To03} when projecting onto the cylinder axis of the nanotubes, with $2(n_1+n_2+\cdots+n_K)$ carbon atoms superimposed onto each other at each point in the 1D point pattern. As a result, the number variance of the MWNTs $\sigma_{N}^2(R)$ can be theoretically determined as
\begin{equation}
\sigma_{N}^2(R) = 4(n_1+n_2+\cdots+n_K)^2 \sigma_{N,S}^2(R),
\end{equation}
where $\sigma_{N,S}^2(R)$ is the number variance of the projected single-scale 1D pattern. On the other hand, as SW defects are introduced into the nanotubes, the behaviors of these MWNTS consisting of purely armchair nanotubes are similar to those MWNTS consisting of purely zigzag nanotubes. For example, we show the computed $\sigma_{N}^2(R)$ for MWNTs consisting of a $(3,3)$-armchair nanotube and a $(5,5)$-armchair nanotube at different defect fractions $p$ in Fig. \ref{fig_3}(b). Importantly, at all investigated $p$, the variance $\sigma_{N}^2(R)$ of these MWNTs fluctuate around certain constants, indicating that these structures are class-I hyperuniform.

Interestingly, MWNTs consisting of both zigzag and armchair nanotubes exhibit different behaviors in their density fluctuations from SWNTs or MWNTs consisting of purely zigzag or armchair nanotubes. For example, as shown in Fig. \ref{fig_3}(c), $\sigma_{N}^2(R)$ of MWNT consisting of a (5,0)-zigzag nanotube and a (5,5)-armchair nanotube exhibit no ``periodicity'' in the window radius $R$ even at low $p$, and appear close to those of a quasi-crystal \cite{To18a}. This is a direct result of the fact that the length of the smallest repeating unit in a defect-free (5,0)-zigzag nanotube and that of the smallest repeating unit in a defect-free (5,5)-armchair nanotube do not have an integer common multiple, and the 1D projections of the two nanotubes collectively onto the cylinder axis are no longer periodic. Here when placing observation window in the process of computing $\sigma_{N}^2(R)$, we restrict the observation window to fall entirely in the range of the (5,5)-armchair nanotube in the axial direction, which is smaller than that of (5,0)-zigzag nanotube. 

The results of ensemble-averaged $S(k)$ of MWNTs consisting of a $(3,0)$-zigzag nanotube and a $(5,0)$-zigzag nanotube, and MWNTs consisting of a $(3,3)$-armchair nanotube and a $(5,5)$-armchair nanotube, are shown in Fig. \ref{fig_2}(d) and Fig. \ref{fig_2}(e), respectively, which all decreases to zero as $k$ goes to zero, regardless of defect concentration $p$. These results further confirm the hyperuniformity of these nanotubes, and are consistent with the results of $\sigma_{N}^2(R)$. We note that due to the lack of periodicity of the simulation box in the axial direction of the MWNT consisting of a (5,0)-zigzag nanotube and a (5,5)-armchair nanotube, $S(k)$ is not well defined for this MWNT, and we do not compute its $S(k)$.

\section{Stability and Density of States of Disordered Hyperuniform Carbon Nanotubes}
We use (10,0) zigzag SWNTs as examples to demonstrate the effect of different concentrations of SW defects on the stability and electronic structure properties of the amorphous SWNTs. Specifically, density functional theory (DFT) \cite{64-Hohenberg,65-Kohn} calculations are performed with the Atomic-orbital Based Ab-initio Computation at UStc (ABACUS) package \cite{10JPCM-Chen,16CMS-Li}.
The norm-conserving pseudopotential \cite{13B-Hamann,15CPC-Schlipf} is employed to describe the ion-electron interactions and the valence electron configuration of C is $2s^{2}2p^{2}$. 
The generalized gradient approximation (GGA) in the form of the Perdew-Burke-Ernzerhof
(PBE) \cite{96L-Perdew} is used for the exchange–correlation functional.
In order to deal with large systems of (10,0) zigzag SWNTs, we employ numerical atomic orbitals (NAO) in the form of double-$\zeta$ plus polarization function (DZP) orbitals as basis sets in our calculations,
whose accuracy and consistency have been verified in previous studies \cite{16CMS-Li,21JNM-Liu,22PCCP-Liu}.
Specifically, we use $2s2p1d$ NAOs for C, 
whose radius cutoff is set to 8 Bohr.
The energy cutoff is set to 60 Ry.
Besides, oeriodic boundary conditions and a single k-point ($\Gamma$ point) are used. 
We employ the Gaussian smearing method with the smearing width of 0.001 Ry.
Structural optimizations are performed with the conjugated gradient method
until forces on each atom are below 0.04 eV/\AA.

\begin{figure}[ht!]
\begin{center}
$\begin{array}{c}\\
\includegraphics[width=0.495\textwidth]{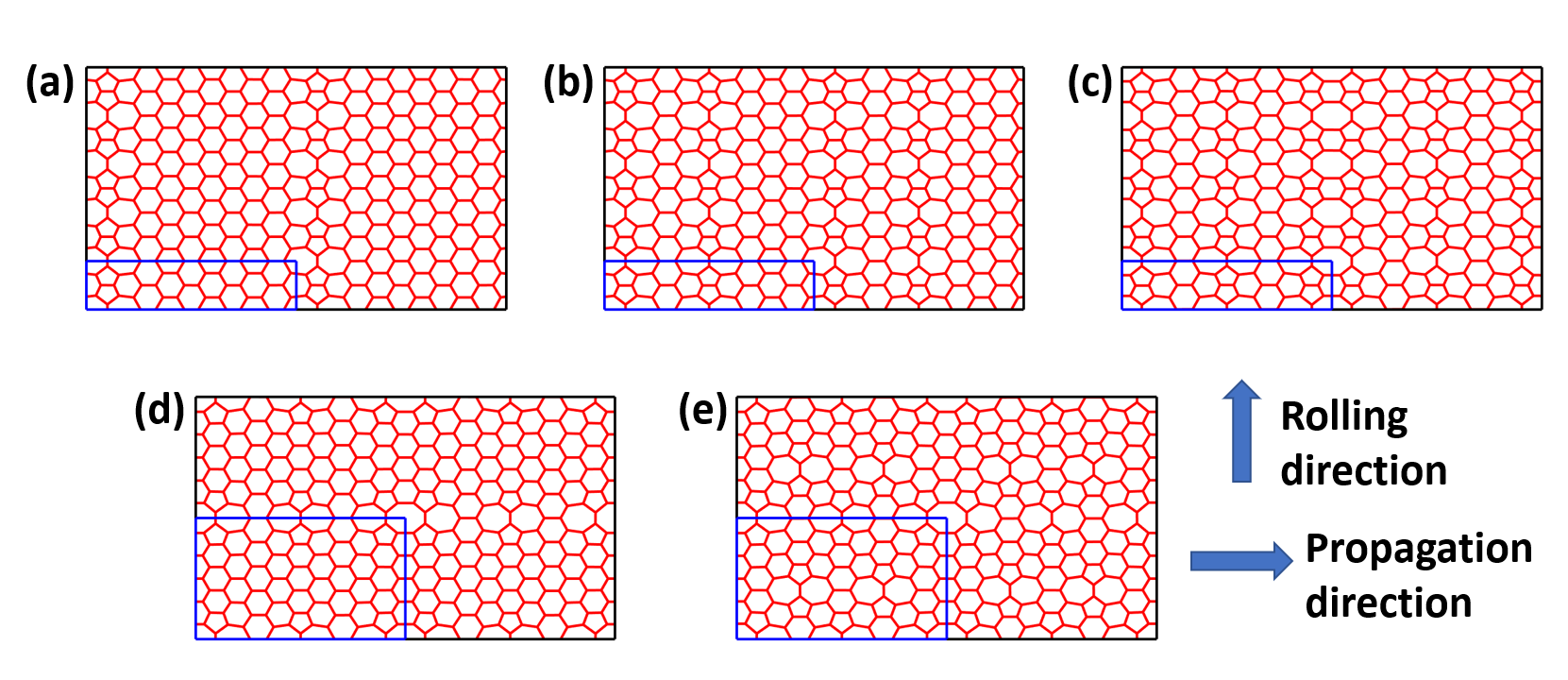} 
\end{array}$
\end{center}
\caption{(Color online) Two classes of defected graphene sheets with periodic distribution of Stone-Wales defects, which are then rolled into (10,0) nanotubes. Their unit cells are shown by the blue boxes, which are replicated five times in the horizontal direction (propagation direction of the nanotubes) before rolling into the nanotubes, and for clear visualization we only show a portion in the horizontal direction. (Top) Defect model I sheets at (a) $p=0.0167$, (b) $p=0.0333$, and (c) $p=0.05$. (Bottom) Defect model II sheets at (d) $p=0.02$ and (e) $p=0.0333$.} \label{fig_4}
\end{figure}

We first demonstrate the stability of DHU nanotubes. To this end, we compare the results of the DHU configurations at different defect concentrations $p$ to those of the defect-free counterpart, as well as two other representative models of nanotubes containing periodically distributed SW defects shown in Fig. \ref{fig_4}. Specifically, in defect model I we introduce SW defects into a unit cell with 40 carbon atoms shown by the blue box in the top row of Fig. \ref{fig_4}, and then replicate the unit cell five times in both horizontal and vertical directions to generate a graphene sheet before rolling up to form (10,0) nanotubes. In defect model II, we introduce SW defects into a unit cell with 100 carbon atoms shown by the blue box in the bottom row of Fig. \ref{fig_4}, and then replicate the unit cell five times in the horizontal direction and two times in the vertical directions to generate a graphene sheet before rolling up to form (10,0) nanotubes. Due to the constraints associated with the construction of defected nanotube models, we only consider three distinct defect concentrations for defect model I SWNTs, and two concentrations for defect model II SWNTs, which we believe are sufficiently representative.

\begin{figure}[ht!]
	\centering
	\includegraphics[width=8cm]{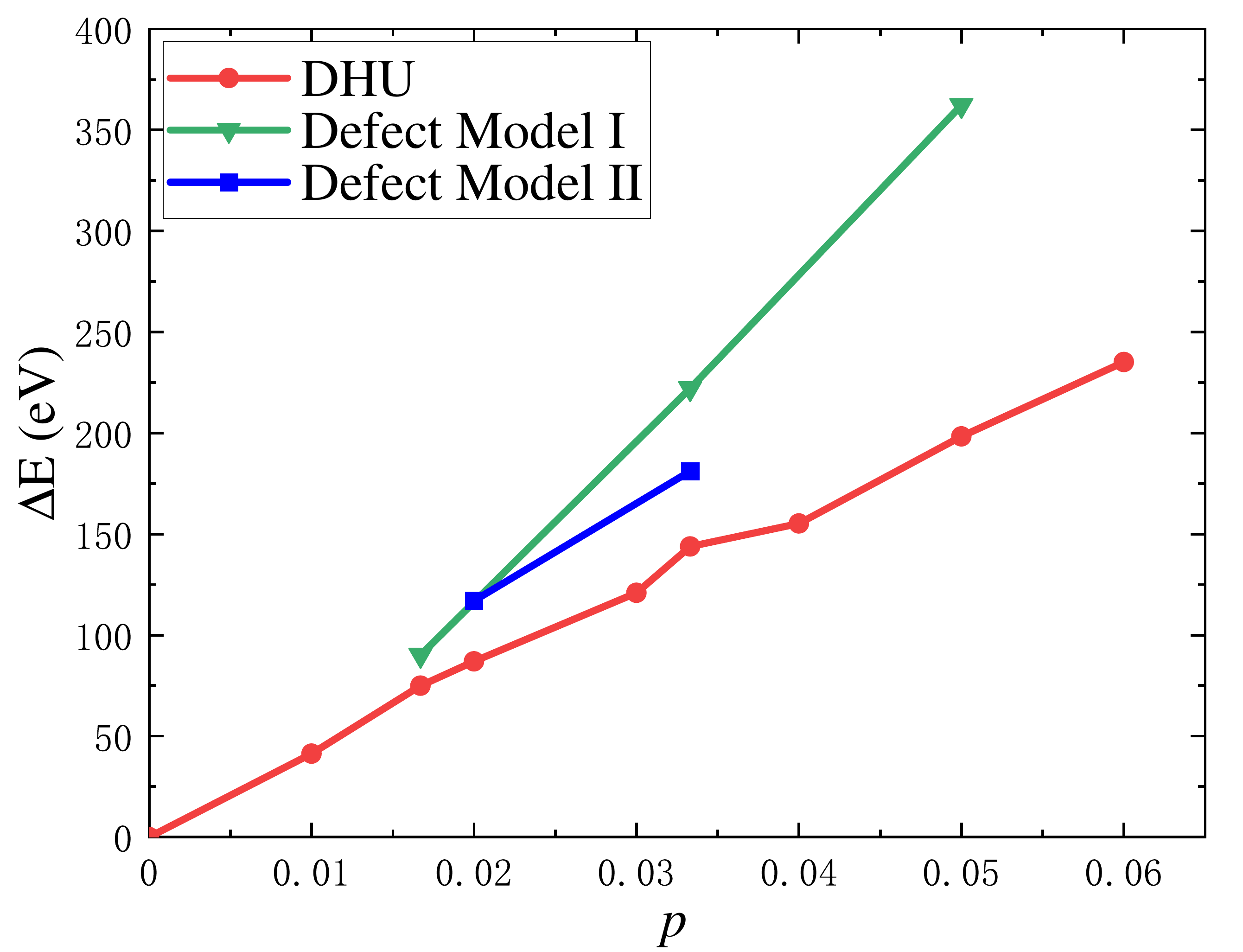}
	\caption{(Color online) Excess energy $\Delta E$ (difference of the total energy between the amorphous system and the corresponding defect-free state) of (10, 0) zigzag DHU nanotubes at different SW defect concentrations $p$, compared to $\Delta E$ of two other representative defect nanotube models with periodically distributed SW defects.
	}
	\label{fig:energy}
\end{figure}

In particular, we calculate the total energy $E$ (i.e., the sum of interaction energy and electronic kinetic energy of many-body systems) of DHU carbon nanotubes at $T = 0~K$ with a wide spectrum of defect concentration $p$ after structural optimizations, as well as the total energy of the two defect models at selected $p$. Fig.~\ref{fig:energy} shows the excess energy $\Delta E$ of the aforementioned systems compared to the defect-free carbon nanotube. Since introducing defects always increases the energy of the system, we have $\Delta E>0$ for all $p>0$. It can be seen that for all three different defected nanotube models, $\Delta E$ increases approximately linearly as the concentration $p$ increases. The DHU nanotubes appear to possess the lowest $\Delta E$ among the three models for the same $p$ values, and possess the smallest slope among the three. This result implies that DHU states are more energetically stable than defect models with periodically distributed SW defects.

\begin{figure*}[ht!]
	\centering
	\includegraphics[width=17cm]{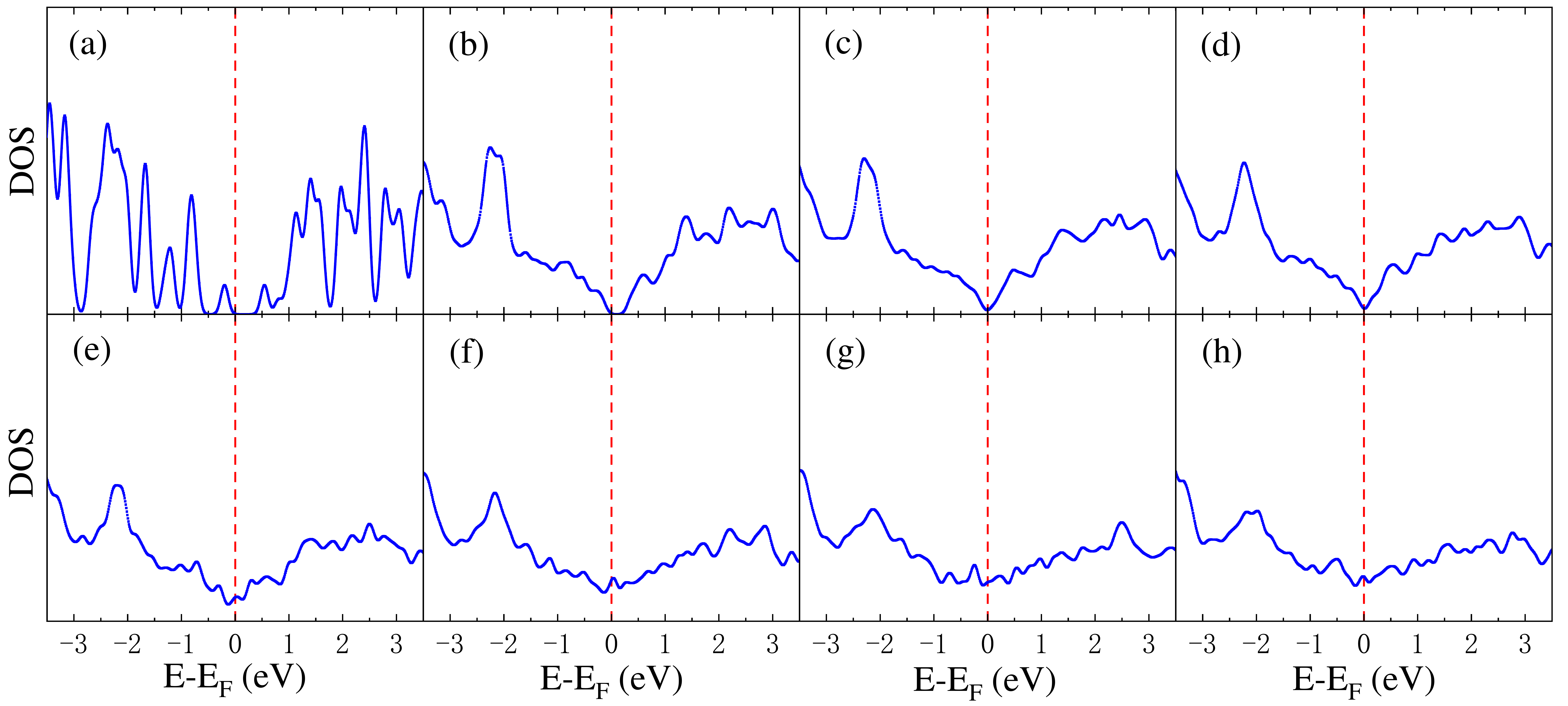}
	\caption{
		(Color online) Density of states (DOS) of (10, 0) zigzag DHU carbon nanotubes computed by PBE functional at defect concentration $p$: (a) 0.00, (b) 0.01, (c) 0.0167, (d) 0.02, (e) 0.03, (f) 0.04, (g) 0.05, (h) 0.06.
	}
    \label{fig:DOS}
\end{figure*}

We further investigate the electronic structures of DHU nanotubes to shed light on the effects of increasing SW defects on the material properties. Fig.~\ref{fig:DOS} show the DOS of (10, 0) zigzag DHU carbon nanotubes computed by PBE functional at different $p$ as well as the DOS of defect-free carbon nanotubes. It is well known \cite{Az10, Pa13} that (10,0) nanotubes possess a well-defined band gap at Fermi level, which is captured in our calculations (see Fig.~\ref{fig:DOS}(a)). Increasing disorder in the DHU system (i.e., increasing the amount of SW defects) results in two observed effects on the computed DOS: (i) closure of the band gap at Fermi level and (ii) broadening and flattening of the DOS, which are consistent with previous computational study of defected nanotube that is semiconducting in its defect-free state \cite{Cr97}. This trend is also similar to the observations in DHU 2D materials \cite{Zh20, Ch21}. Specifically, the band gap is closed at $p = 0.0167$, suggesting the presence of the semiconductor-to-metal transition around this defect concentration. Moreover, as $p$ increases the DOS becomes more and more extended, converging to a metallic characteristic. We note although the exact values of the band gaps could be underestimated by PBE functional compared with the actual values, the qualitative behavior of the DOS in DHU nanotubes suggests a semiconductor-to-metal transition should be robust.

\section{Conclusions and Discussion}

In this work, we generalized the concept of hyperuniformity to structurally characterize quasi-one-dimensional materials, which in the context of hyperuniformity can be viewed as one-dimensional projections of higher-dimensional structures along the axial direction. As a proof of concept, we systematically investigated the density fluctuations across length scales in amorphous carbon nanotubes containing different amounts of Stone-Wales topological defects, which can be constructed by rolling up defected graphene sheets. We demonstrated that all amorphous nanotubes containing SW defects studied here are hyperuniform, i.e., the normalized infinite-wavelength density fluctuations are completely suppressed, regardless of the diameter, rolling axis, number of rolling sheets, and defect fraction of the nanotubes. Disordered hyperuniformity (DHU) is recently discovered exotic state of matter. Using DFT simulations, we also showed that these amorphous carbon nanotubes with randomly distributed Stone-Wales defects are energetically more stable than their ordered counterparts with periodically distributed Stone-Wales defects.  We also demonstrated that the electronic bandgap closes for a semiconducting zigzag nanotube as Stone Wales are randomly introduced into the carbon nanotubes by determining the density of states near the Fermi level. Our structural study of amorphous nanotubes strengthens our fundamental understanding of these quasi-1D materials, and suggests possible exotic physical properties, as endowed by unique disordered hyperuniformity feature.

\begin{figure}[ht!]
\begin{center}
$\begin{array}{c}\\
\includegraphics[width=0.45\textwidth]{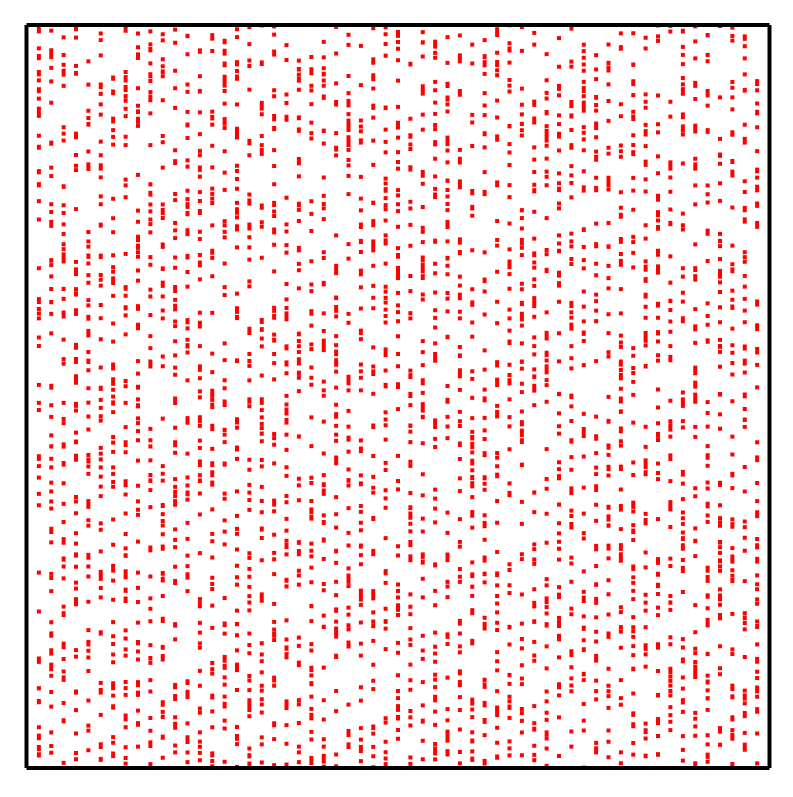} 
\end{array}$
\end{center}
\caption{(Color online) An anisotropic point pattern where the points are placed randomly in the vertical direction, but are constrained to discrete equally-distributed lattice sites in the horizontal direction with each sites having the same number of points. The projection of this point pattern is hyperuniform when projected along the horizontal direction, and nonhypeurniform when projected along the vertical direction.} \label{fig_7}
\end{figure}

Our findings on the effect of the projection operation on the hyperuniformity property of the graphene sheets may also shed light on our understanding of the general effect of dimensionality reduction on the preservation of (non)hyperuniformity, as projection is a common type of dimensionality reduction operation. While we conjecture that projection should preserve (non)hyperuniformity of isotropic higher-dimensional structures since each dimension contributes equally to the density fluctuations in the isotropic cases, this is definitely not always the case for anisotropic structures. This can be seen from the following example of anisotropic point patterns, where the points are placed randomly in the vertical direction, but are constrained to discrete equally-distributed lattice sites in the horizontal direction with each sites having the same number of points, as shown in Fig. \ref{fig_7}. The projection of this two-dimensional point pattern along the vertical direction is a one-dimensional Poisson point pattern, which is known to be nonhyperuniform \cite{To03}. On the other hand, the projection of this two-dimensional point pattern along the horizontal direction is a single-scale lattice in one dimension, which is known to be class-I hyperuniform \cite{To03}. This example shows that the projection of an anisotropic point could be hyperuniform when projected one direction, but nonhyperuniform when projected along another direction.

\begin{acknowledgments}
Y.L and M.C. were supported by the National Science Foundation of China under Grant Nos. 12122401 and 12074007. Parts of the numerical simulations were performed on the High Performance Computing Platform of CAPT. H. Z. thanks the start-up funds from ASU. This research also used computational resources of the Agave Research Computer Cluster of ASU and the Texas Advanced Computing Center under Contract No. TG-DMR170070. 
\end{acknowledgments}


\end{document}